\documentclass[%
 reprint,
 amsmath,amssymb,
 aps,
]{revtex4-1}

\usepackage{graphicx}% Include figure files
\usepackage{dcolumn}% Align table columns on decimal point
\usepackage{bm}% bold math
\begin{document}

\preprint{APS/123-QED}

\title{The heterodyne sensing system for the ALPS II search\\ for sub-eV weakly interacting particles}% Force line breaks with \\

\author{Ayman Hallal}
\author{Giuseppe Messineo}
\author{Mauricio Diaz Ortiz}
\author{Joseph Gleason}
\author{Harold Hollis}
\author{D.~B.~Tanner}
\author{Guido Mueller}
 \email{mueller@phys.ufl.edu}
\affiliation{%
 Deparment of Physics, University of Florida\\
Gainesville, Florida, U.S.A.
}%

\author{Aaron Spector}
\affiliation{
DESY\\
 Hamburg, Germany
}

\date{\today}% It is always \today, today,

\begin{abstract}
 ALPS II, the Any Light Particle Search, is a second-generation Light Shining through a Wall experiment that hunts for axion-like particles. The experiment is currently transitioning from the design and construction phase to the commissioning phase, with science runs expected to start in 2021. ALPS II plans to use two different sensing schemes to confirm the potential detection of axion-like particles or to verify an upper limit on their coupling strength to two photons of $g_{a\gamma\gamma}\leq2\times10^{-11}\text{GeV}^{-1}$.
This paper discusses a heterodyne sensing scheme (HET) which will be the first  scheme deployed to detect the regenerated light. It presents critical details of the optical layout, the length and alignment sensing scheme, design features to minimize spurious signals from stray light, as well as several control and veto channels specific to HET which are needed to commission and operate the instrument and to calibrate the detector sensitivity.
\end{abstract}

\maketitle
\let\clearpage\relax

\section{Introduction \label{sec:Introduction}}

\begin{figure*}[t]
\begin{centering}
\includegraphics[width=16cm]{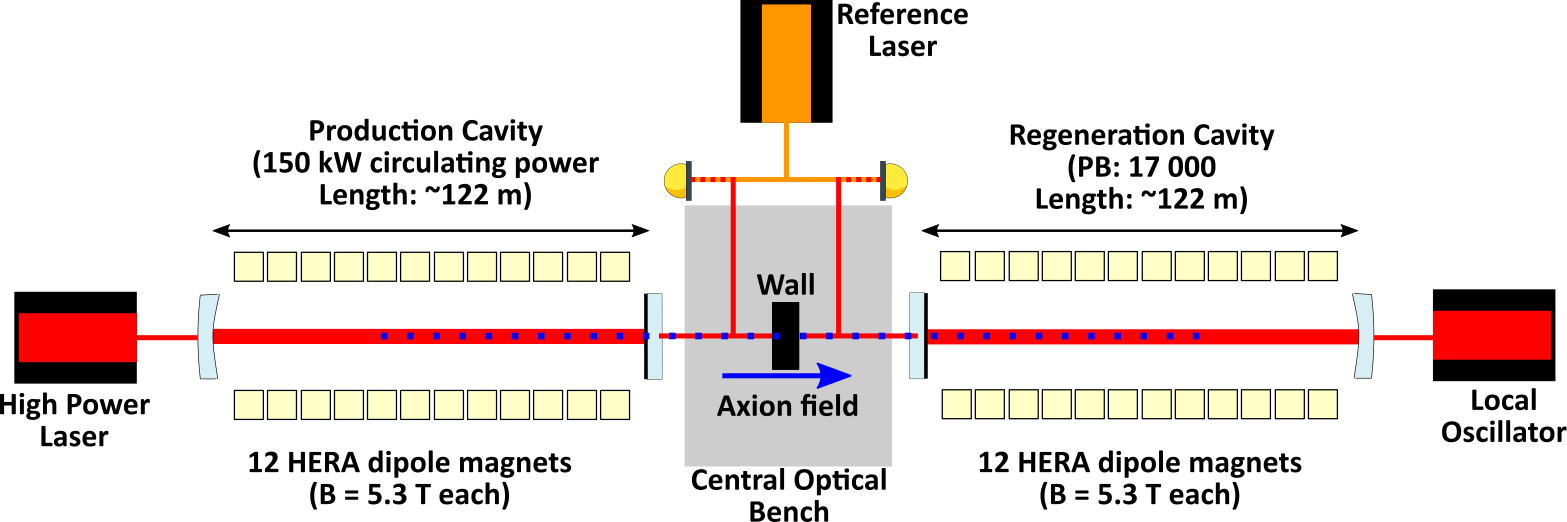}
\par\end{centering}
\caption{\label{fig: Experimental layout of ALPS II:}Experimental layout of ALPS~II. A high power laser is injected into the 122~m long production cavity on the left. 150~kW of intra-cavity light propagates back and forth inside a string of 5.3~T HERA dipole magnets. A small fraction of this field is turned into an axion field which propagates through the wall into an identical string of twelve HERA magnets. The magnetic field turns a small fraction of the axion field back into light which is resonantly enhanced by the regeneration cavity. A local oscillator laser is used to (a) sense the length and alignment of the regeneration cavity and (b) act as a local oscillator to detect the regenerated field. A reference laser is used to maintain coherence between all fields.}
\end{figure*}

%%Removed LSW. It was not used again.
The Any Light Particle Search, currently under construction at DESY in Hamburg, Germany, will be the largest and most sensitive light-shining-through-a-wall experiment ever built ~\cite{ALPS-opticspaper}. ALPS~II will search for axion-like pseudo-scalar and scalar particles as well as hidden sector photons with masses in the sub-meV/$c^2$ range. (In the remainder of the paper, we refer to all these particles as axions or axion-like particles.) The fundamental parts of ALPS~II are shown in Figure~\ref{fig: Experimental layout of ALPS II:}. The apparatus consists of two strings of twelve straightened 8.8~m long, 5.3~T HERA magnets which are separated by a light-tight wall~\cite{Clemens2020}. Inside the magnet string on the left side of the wall, some of the photons from a high power laser will be converted into axion-like particles. These axion-like particles pass through the wall unimpeded and enter the second string, where some transform back into photons indistinguishable from the original laser photons. The detection of photons on the dark side of the wall identical to photons on the bright side of the wall would confirm the existence of axion-like particles~\cite{van_bibber_proposed_1987}. To increase the number of generated axion-like particles, the laser field inside the first magnet string is resonantly enhanced using a 124~m long optical cavity, named the production cavity. A similar regeneration cavity is also used on the other side of the wall to resonantly enhance the number of regenerated photons~\cite{Hoogeveen90, robilliard_no_2007, chou_search_2008, ehret_new_2010}. Initially, ALPS~II will employ a heterodyne sensing scheme (HET)~\cite{mueller_detailed_2009, bush_coherent_2019} described in detail in this paper. This search will be followed later by a transition edge sensor (TES) based sensing scheme to verify and confirm the HET results.

The basic properties of and methods used in ALPS~II are described in a second paper ~\cite{ALPS-opticspaper}. That paper focuses on the primary requirements to achieve our targeted sensitivity and discusses how the optical system is designed to meet these requirements while providing as robust of an experimental setup as possible. Our paper describes the final design and implementation of the HET. This includes an exploration of the system architecture that provides sensing signals in addition to those of the optical system, technical details related to the hardware and data processing, and the methods that will be used to verify the efficacy of the measurement.

The heterodyne detection system uses the coherence between the axion production and photon regeneration processes. It can detect very weak photon fields, at the shot-noise limit as demonstrated in~\cite{bush_coherent_2019}. By optically mixing the regenerated field with a much stronger optical field, called the local oscillator (LO) field, we generate a beat signal varying in time at the frequency difference of the two fields. This beat signal is a measure of the amplitude and the phase of the field regenerated in the second magnet string. Performing an in-phase and quadrature (I/Q) demodulation at the frequency difference between the two optical fields and integrating both quadratures over a long measurement time, one can construct a quantity proportional to the regenerated photon rate. The detection method is essentially a running single-bin discrete Fourier transformation where the bandwidth of the single bin progressively decreases as the inverse of the integration time. This heterodyne sensing scheme is compatible with the top-level requirements for the ALPS~II science run listed in Table~1 in~\cite{ALPS-opticspaper}. %% TLR not used again.

%%Chapter-> section. Only books, theses have chapters.
%Note that throughout the paper we will use the admittedly somewhat clumsy project nomenclature for specific components to be consistent with technical notes and future publications.

%\onecolumngrid

%\twocolumngrid

The following section will give an overview of the design of ALPS~II using the heterodyne sensing system. One of the key aspects at this stage is the management of the laser power at the different locations in the experiment. This is a critical balancing act between maintaining the signal to noise ratio for several essential signals and limiting the stray light background as much as possible. The third section introduces the heterodyne sensing scheme and the associated stringent requirements on the stability of the interferometric setup. This section also discusses the expected leading limitations as well as some of the more important countermeasures that are installed to allow us to achieve our sensitivity goal. The fourth section focuses on the alignment process and the in-situ monitoring of the alignment, the calibration of a veto signal and the validation of the coupling of the axion field into the regeneration cavity; a critical aspect of the initial commissioning of the experiment that would otherwise either prevent a detection or limit our ability to place an upper limit on the axion to two-photon coupling coefficient. Stray light mitigation techniques will be discussed in the fifth section while the last section provides a summary and conclusions. 

\section{Overview \label{sec:Overview}}

\begin{figure*}[t]
\begin{centering}
\includegraphics[width=14cm]{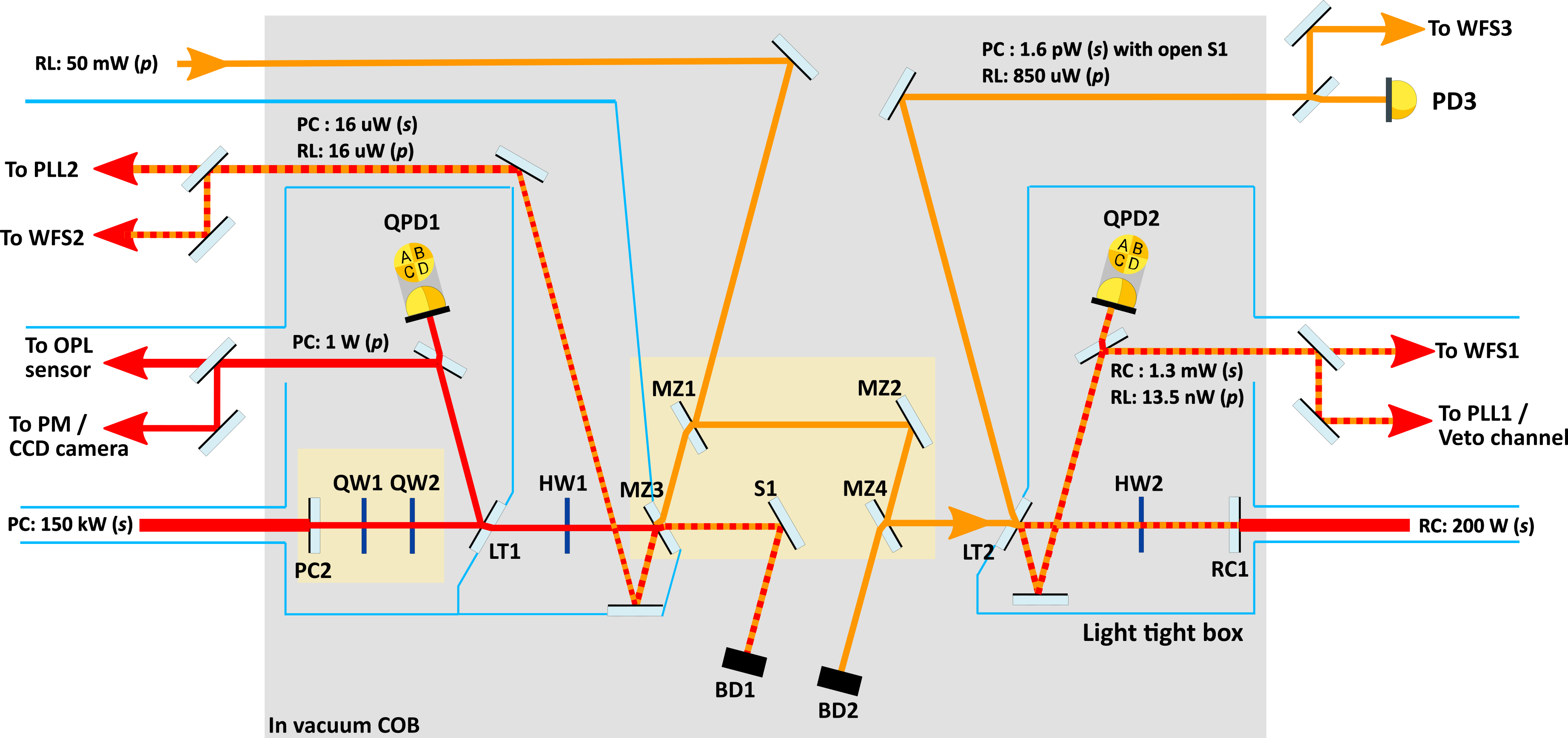}
\par\end{centering}
\caption{\label{fig:Optics_RL_COB} Optical layout of the central optical bench (COB). HW: Half-Wave plate, MZ: partially transmissive mirror of the Mach-Zehnder-like interferometer, LT: partially transmissive mirror of the Light-Tight box, PC: Production Cavity, OPL: Optical Path Length, CCD camera: monitors spatial mode, PD: Photodetector, PLL: Phase-Locked Loop, QW: Quarter-Wave plate, QPD: Quadrant Photodetector, RL: Reference Laser, RC: Regeneration Cavity, S1: Shutter, WFS: Wavefront Sensor, BD: Beam Dump,  \textit{s}: vertical polarization direction, \textit{p}: horizontal polarization direction. Note that the polarization in both cavities can be changed while the waveplates on the COB but also in the end stations allow to maintain the polarizations everywhere else. The power levels of the three laser fields are estimated based on known mirror reflectivities and transmissivities.}
\end{figure*}

A conceptual layout of the experiment is shown in Figure~\ref{fig: Experimental layout of ALPS II:}. The high power laser (HPL) will be injected from the left end station into the production cavity (PC) while the local oscillator (LO) laser will be injected from the right end station into the regeneration cavity (RC). The frequency of each laser will be locked to its respective cavity using the Pound-Drever-Hall (PDH) technique. Figure~\ref{fig:Optics_RL_COB} shows the area of the central optical bench (COB) and its beam paths and sensors. The COB is the centerpiece of the experiment. It is located inside the vacuum chamber and connected to the two magnet strings via vacuum bellows. The transverse positions of the cavity eigenmodes will be controlled using the position sensors QPD1 and QPD2 on the COB. The central area, outside the COB, contains other sensors. As shown, most of the PC transmitted light will be directed out of the vacuum chamber where it is used to measure the transmitted power, to monitor intensity distribution of the laser beam, and to sense the optical path length changes in the PC cavity mirror substrate PC2. (See section \ref{OPL_Sensing} for details). 

%%ccd not mentioned again. 

A third laser, called the reference laser (RL), is located in the central area and is used as an intermediate reference. It transfers information about the actual resonance frequency from the RC to the PC and monitors the relative alignment of the transmitted cavity fields with respect to each other. The use of the RL avoids the need for a direct connection between the PC and RC output fields, reducing possible light contamination from the PC to the RC. The RL will be injected from the PC side of the COB into the Mach-Zehnder-like interferometer (MZ) formed by the four mirrors MZ1, MZ2, MZ3, and MZ4. These mirrors are mounted on  an ultra-low expansion (ULE) base plate, described in section \ref{MZ}. The ULE plate  is in turn mounted on a large aluminum breadboard inside the COB vacuum tank. One part of the RL beam is combined with a small fraction of the PC transmitted power at MZ3 and directed through a vacuum viewport on the left side. This superposition is used to phase-lock the PC transmitted field to the RL (PLL2) and to monitor the alignment between the two fields using a wavefront sensor (WFS2). The second part of the RL light is mostly reflected at LT2 and then directed through a vacuum viewport on the right side of the COB. The field then goes to a pair of quadrant detectors (WFS3) and a single element detector PD3. With open shutter (S1), these detectors monitor the amplitude and phase of the PC transmitted light as well as the relative alignment with respect to the RL. With closed shutter, PD3 monitors stray light at the HPL frequency in the spatial mode of the RL while the DC signals of the WFS3 detector provide auxiliary signals to monitor the alignment of the RL through the COB.

A small fraction of RL passes through LT2 and combines with the RC transmitted light at RC1. This superposition is directed out of the vacuum chamber. Most of the light is used to generate the signal for the phase-locked loop (PLL1) of the RL to the RC-transmitted field while a smaller fraction is directed towards a   wavefront sensor  (WFS1) that generates the error signal for the alignment of the RL relative to the RC transmitted field. The signal from PLL1 is also used to search for spurious signals at $\Omega_{sig}$. These signals could be produced by regenerated photons as well as by unwanted stray light. This procedure is described in section \ref{sec:Stray}. The nominal values for the PLL frequencies and the PDH/WFS modulation frequencies are shown in Table~\ref{frequencies}. We will also use another modulation frequency to measure and, if needed, to stabilize the free spectral range (FSR) of the RC. This frequency is further modulated by an audio frequency to generate the required error similar to what is done in Ref.~\cite{thorpe_laser_2008}.

\begin{table}
\begin{centering}
\begin{tabular}{|c|c|c|c|}
\hline 
Frequency & Symbol & Value & Comment \tabularnewline
\hline 
Signal frequency & $\Omega_{Sig}$ & 15.6~MHz & $13\times \rm{FSR}$\tabularnewline
\hline 
PLL RL/RC-Trans & $\Omega_1$ & 9.0~MHz  & $7.5\times \rm{FSR} + 133~\rm{Hz}$ \tabularnewline
\hline 
PLL RL/PC-Trans & $\Omega_2$ & 6.6~MHz & $5.5\times \rm{FSR} - 133~\rm{Hz}$ \tabularnewline
\hline 
PDH/WFS at HPL & $f_{1}$& 8.0~MHz &  \tabularnewline
\hline 
PDH/WFS at LO & $f_{2}$ & 9.2~MHz &  \tabularnewline
\hline 
FSR sensing & $f_{FSR}$ & 12 MHz & $10.0\times \rm{FSR}$ \tabularnewline
\hline 
\end{tabular}
\par\end{centering}
\caption{\label{frequencies} Frequency plan for the heterodyne sensing scheme. The signal frequency, $\Omega_{Sig}$, is the difference between the frequencies of the PC and LO lasers. The exact values of the two PLL frequencies, $\Omega_1$ and $\Omega_2$, are not critical as long as the RL is not resonant in the RC or PC. However,  $\Omega_1$ and $\Omega_2$ have to add up to the signal frequency $\Omega_{Sig}$. For all other frequencies: none of the sum or difference frequencies between the fundamental and higher harmonics should combine to the signal frequency.}
\end{table}

\begin{figure}
\begin{centering}
\includegraphics[width=8cm]{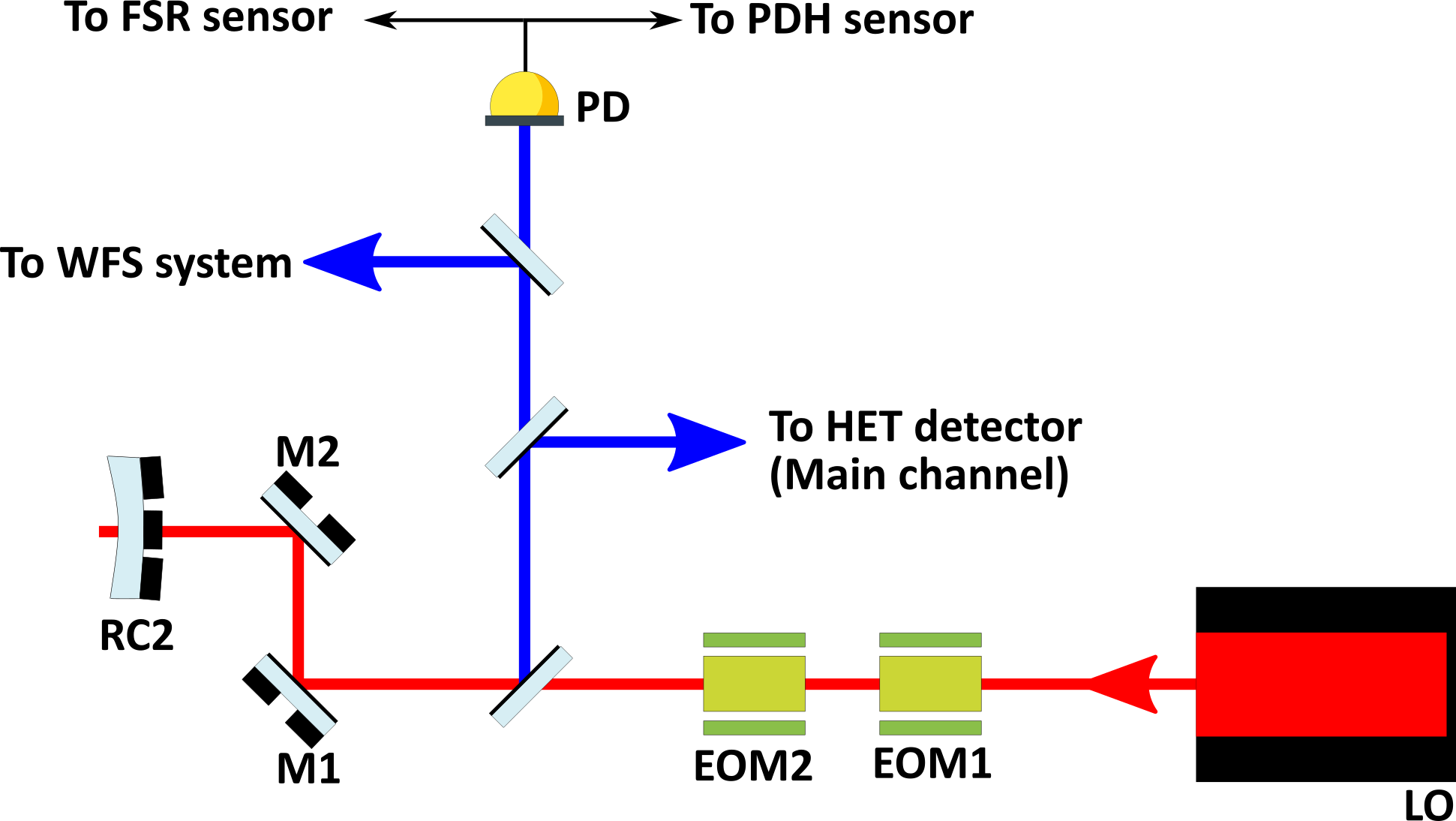}
\par\end{centering}
\caption{\label{fig:Optics_ET}Optical layout of the injection system for the local oscillator laser.  EOM: Electro-Optical Modulator, FSR: Free Spectral Range, HET: Heterodyne sensor, LO: Local Oscillator, M: high reflective Mirror, RC: Regeneration Cavity, PD: Photodetector, PDH: Pound-Drever-Hall, WFS: Wavefront Sensor.}
\end{figure}

Figure~\ref{fig:Optics_ET} shows a sketch of the injection system for the local oscillator laser into the RC. These optics are located on the right end table. The LO is a diode-pumped monolithic Nd:YAG laser emitting at 1064~nm. Two electro-optical phase modulators in series, EOM1 and EOM2, are used to generate phase modulation sidebands for the PDH and the FSR sensing scheme. Most of the reflected field will be directed to a dedicated HET detector to measure the beat signal between the regenerated photons and the LO. Smaller fractions are used for the PDH sensor, the FSR sensor, and the pair of quadrant detectors that form the WFS system. The alignment signals are used in a servo system to maintain alignment of the LO into the RC by acting on the two piezo-controlled alignment mirrors M1 and M2. The injection system of the high power laser on the left end table is similar to the LO system, except that the HET detector is not required~\cite{ALPS-opticspaper}. 

\subsection{Power management}
As mentioned earlier, ALPS~II will need careful management of the power of the laser fields in the different areas of the experiment. These fields must form adequate beat signals to achieve the required detection efficiency and to minimize signal loss due to residual phase noise and cavity and laser beam misalignments~\cite{ALPS-opticspaper}.
Specifically, the power of the local oscillator laser in the beat notes must be between 0.5 and 10\,mW wherever shot-noise limited detection is needed, such as in the HET detector and the veto channel. Also, the power in the beat note must be sufficient to allow phase-locks between two laser fields  with less than 0.01~rad phase noise and enable measurements of tilts between interfering wavefronts with $\theta_{Div}/100$ sensitivity, where $\theta_{Div}$ is the divergence angle of the cavity eigenmode. On the other hand, the design must minimize the amount of stray light entering the RC which could lead to a false signal indistinguishable from the regenerated signal. 

Table~\ref{mirrors} shows the power transmissivities for both polarizations of all critical mirrors on the optical bench. This design leads to the estimated power levels shown in Figure~\ref{fig:Optics_RL_COB} at the beam splitters where the key beat signals are formed. Note that the optical axes of the two quarter-wave plates QW1 and QW2 are aligned and rotate the polarization of the \textit{s}-polarized PC transmitted field to \textit{p}-polarization to increase the transmission through LT1. The role of these two quarter-wave plates will be further discussed in section~\ref{fig:OPL-sensing}. The half-wave plate HW1 rotates the polarization back to \textit{s} after LT1 to take advantage of the lower transmissivities of the MZ mirrors in \textit{s}-polarization. This design helps to reduce stray light reaching the RC.
%Note that the two quarter-wave plates QW1 and QW2 with their optical axes aligned are placed after the PC to rotate the polarization of the s-polarized PC transmitted field to p-polarization and increase the transmission through LT1. 
\begin{table}

\begin{centering}
\begin{tabular}{|c|c|c|c|c|}
\hline 
Mirror & AoI & $T_{s}$ & $T_{p}$ & Comment\tabularnewline
\hline 
\hline 
PC1 & $0^{\circ}$ & 107~ppm & 107~ppm & Vendor data \tabularnewline
\hline 
PC2 & $0^{\circ}$ & 6.7~ppm & 6.7~ppm & Vendor data \tabularnewline
\hline 
RC2 & $0^{\circ}$ & 107~ppm & 107~ppm & Vendor data \tabularnewline
\hline 
RC1 & $0^{\circ}$ & 6.7~ppm & 6.7~ppm & Vendor data  \tabularnewline
\hline 
LT1 & $35^{\circ}$ & 0.5~ppm & 16~ppm & Design value\tabularnewline
\hline 
LT2 & $35^{\circ}$ & 0.5~ppm & 16~ppm & Design value \tabularnewline
\hline 
MZ2 & $35^{\circ}$ & 0.5~ppm & 16~ppm & Design value\tabularnewline
\hline 
MZ1 & $35^{\circ}$ & 0.943 & 0.983 & Uncoated glass\tabularnewline
\hline 
MZ3 & $35^{\circ}$ & 9.6~ppm & 322~ppm & Measured\tabularnewline
\hline 
MZ4 & $35^{\circ}$ & 9.6~ppm & 322~ppm & Measured\tabularnewline
\hline 
\end{tabular}
\par\end{centering}
\caption{\label{mirrors} Transmissivities  of mirrors and beamsplitters critical to managing the laser power inside ALPS~II. AoI: Angle of incidence.}

\end{table}

%Based on surface roughness measurements of the cavity mirrors, we expect scatter losses inside the cavities between 40\,ppm and 70\,ppm \cite{ALPS-opticspaper} resulting in an over-coupled RC seen from the end station. The ratio between the reflected and the transmitted power will be between 0.7 and 2.7. The injected LO power as well as the modulation indices will be optimized to guarantee shot noise limited detection by PPL1 near the central tank and by the HET detector in the end station. 

\section{Signal Demodulation and phase tracking\label{sec:Demod}}

% Should start with a requirement of what we need to achieve here. 

% subsections: Review double demodulation scheme (Matlab last stage),  synchronization of Mokus. 
% Allows to also demodulate off resonance to monitor shot noise and to switch demodulation phases for stray light suppression.

The HET scheme is based on the idea that the regenerated field and
the local oscillator (LO) field inside the RC create a beat signal
with angular frequency $\Omega_{Sig}$ and, if scaled to photon rates,
an amplitude of $S_{0}=2\sqrt{n_{LO}n_{S}}$ where $n_{LO}$ and $n_{S}$
are the photon rates of the two fields. This beat signal is measured
with photodiodes on both sides of the RC, and the resulting electronic
signal is digitized and then multiplied within a field-programmable
gate array (FPGA) by numerically generated sine and cosine signals
at angular frequency $\left(\Omega_{Sig}+2\pi\Delta f\right)$ where
$\Delta f$ will be of the order of 2.4\,Hz. The resulting I and
Q data streams are low-pass filtered with cascaded integrator-comb
(CIC) filters and then downsampled to approximately 20 samples per
second. These data streams are then I/Q demodulated again with $\Delta f$
in real-time to form four data streams: 
\begin{eqnarray}
II(t) & = & \int_{0}^{T}\left<S_{0}\cos\left(\Omega_{Sig}t+\phi\right)\cdot\cos\big(\left(\Omega_{Sig}+2\pi\Delta f\right)t\big)\:\right>_{\tau}\nonumber \\
 & \quad & \times\cos\left(2\pi\Delta ft+\theta\right)dt=\frac{\sqrt{n_{LO}n_{S}}}{2}T\cos\left(\Delta\theta\right),\\
IQ(t) & = & \int_{0}^{T}\left<S_{0}\cos\left(\Omega_{Sig}t+\phi\right)\cdot\sin\big(\left(\Omega_{Sig}+2\pi\Delta f\right)t\big)\:\right>_{\tau}\nonumber \\
 & \quad & \times\cos\left(2\pi\Delta ft+\theta\right)dt=\frac{\sqrt{n_{LO}n_{S}}}{2}T\sin\left(\Delta\theta\right),\\
QI(t) & = & \int_{0}^{T}\left<S_{0}\cos\left(\Omega_{Sig}t+\phi\right)\cdot\cos\big(\left(\Omega_{Sig}+2\pi\Delta f\right)t\big)\:\right>_{\tau}\nonumber \\
 & \quad & \times\sin\left(2\pi\Delta ft+\theta\right)dt\propto=-\frac{\sqrt{n_{LO}n_{S}}}{2}T\sin\left(\Delta\theta\right),\\
QQ(t) & = & \int_{0}^{T}\left<S_{0}\cos\left(\Omega_{Sig}t+\phi\right)\cdot\sin\big(\left(\Omega_{Sig}+2\pi\Delta f\right)t\big)\:\right>_{\tau}\nonumber \\
 & \quad & \times\sin\left(2\pi\Delta ft+\theta\right)dt=\frac{\sqrt{n_{LO}n_{S}}}{2}T\cos\left(\Delta\theta\right),
\end{eqnarray}
where the CIC filter is simulated by an averaging $\left<..\right>_{\tau}$
over some time $\tau$ on the order of 50~ms. Here, $\phi$ is the
unknown and ideally time-independent phase of the beat signal while
$\Delta\theta=\theta-\phi$ is the difference between the signal and
the demodulation phase. These four signals can be combined, 
\begin{equation}
S(T)=\sqrt{\left(II+QQ\right)^{2}+\left(IQ-QI\right)^{2}}= \sqrt{n_{LO}n_{S}}T,
\end{equation}
to calculate the amplitude of the beat signal independent of the signal
phase which increases linearly with the measurement time $T$.

The standard deviation in each of the quadratures is limited by shot noise and proportional to the root of the number
of detected photons:
\[
\sigma_{II}=\sqrt{n_{LO}T}\frac{1}{4}=\sigma_{QQ}=\sigma_{IQ}=\sigma_{QI}
\]
These uncorrelated noise contributions add quadratically:
\[
\sigma_{S}=\sqrt{4\sigma_{II}^{2}}=\sqrt{n_{LO}T}
\]
and increase over time with $\sqrt{T}$. The signal to noise ratio:
\[
SNR=\sqrt{n_{S}T}=\sqrt{N_{S}}
\]
is typical for heterodyne detections with an unknown signal phase.
A more detailed discussion of shot noise and the detection statistics can be found in  \cite{bush_coherent_2019, Bush-dissertation}.
For a detection with a $5\sigma$ confidence level, 29 regenerated
photons need to be detected.

Better performance can be achieved by setting $\theta=\phi$ and then
calculating the signal via 
\begin{equation}
S=\left(II+QQ\right)\,,
\end{equation}
which reduces the shot noise by a factor $\sqrt{2}$ and only 14 photons
are needed for a 5-sigma detection. As the signal phase is a priori
not known, we plan to use 18 different demodulation phases ranging
from 0 to $85^{\circ}$ separated by $5^{\circ}$ for the \textit{I}
and \textit{Q} demodulation simultaneously and then form $S=\left(II+QQ\right)$
and $S=\left(IQ-QI\right)$ to optimize the search.

Both techniques are sensitive to changes in the phase $\Delta\theta$.
Small fluctuations in $\Delta\theta$ will move energy into other
frequency bins leading to signal losses proportional to $\Delta\theta_{rms}^{2}$.
Larger drifts shift energy into the other quadrature and changes by
$180^{\circ}$ will invert the signal leading to its cancellation
during the integration.

Systematic and stochastic changes in the signal phase with respect
to the demodulation phase can be caused by:
\begin{itemize}
\item Differential clock noise and RF noise. 
\item Interferometer noise: 
\begin{itemize}
\item Geometric path length changes. 
\item Index of refraction changes in optical substrates on the COB due to
environmental temperature changes. 
\item Index of refraction changes in the PC cavity mirror substrate due
to laser heating. 
\end{itemize}
\end{itemize}
Our goal is to keep $\Delta\phi_{rms}<0.2~{\rm {rad}}$ over the integration
time of the experiment. The next subsections describe our plans to
achieve this ambitious goal.

\subsection{Synchronization of the RF signals}

The expected integration time for the experiment is 20 days or 1.7
million seconds. Occasional loss of lock and other interruptions may
occur. The demodulation performs essentially a single-bin Fourier
transformation of a signal at $\Omega_{Sig}$ with a frequency resolution
of better than $1\,\mu{\rm {Hz}}$. The required phase stability of
the demodulation process requires that 
\begin{equation}
\Omega_{Sig}t=\Omega_{1}t+\Omega_{2}t,
\end{equation}
for the duration of the experiment, where $\Omega_{1}$ and $\Omega_{2}$
are respectively the frequencies used in PLL1 and PLL2 as shown in
Table~\ref{frequencies}. A $1.2\,{\rm {GHz}}$ master oscillator
generates local 10\,MHz clock signals for distribution inside the
central and the end stations. Each of the three frequencies will be
generated by a numerically controlled oscillator (NCO) relative to
the 10\,MHz clock signal. The accuracy of the frequency values set
by the NCOs depends on their architecture and here specifically on
the resolution of the frequency control word and the bit depth of
its phase accumulator. Both have to match in each of the three NCOs
to ensure that $\Omega_{1}+\Omega_{2}=\Omega_{Sig}$ with nHz accuracy.
Furthermore, we also require that the residual differential phase
noise between the NCOs is below 0.1~rad rms during the integration
time of the experiment.

We use NCOs inside of identical instruments (Moku:Labs from Liquid
Instruments~\cite{Moku}) to demodulate the two PLL signals as well
as all HET signals. We tested that this setup indeed provided proper
frequency matching and that the residual phase noise is below 0.01~rad
rms over several days of integration time.

As mentioned above, the data will likely not be taken continuously
but interrupted by the occasional loss of lock resulting in the need
to stitch the data together. The clocks, Moku:Labs, and the front
end computer will be connected to a UPS to ensure continuous data
taking of the HET and veto channels. Data flags will be used to distinguish
between useful search data when all systems work properly and the
shutter is closed, useful calibration data when all systems work properly
and the shutter is open, and useless data when at least one of the
subsystems is not working properly.

\subsection{Interferometer noise}

The axion field propagates from the PC to the RC unimpeded and unchanged
by any substrate in its path. In other words, the phase difference
between the axion field inside the PC and inside the RC depends only
on the geometric distance between the cavities. The role of the MZ
interferometer on the COB is to track all geometric distance changes
between the high reflecting (HR) surfaces of the PC and RC mirrors
and turn them into phase changes of the PC transmitted light, using
the two sequential phase-locked loops PLL1 and PLL2, and the PDH servo
system. These systems maintain the necessary phase coherence between
the regenerated field (that originates from the PC circulating field)
and the RC resonating field, required for a correct signal demodulation.
However, the need to completely block the PC transmitted light from
reaching the RC complicates the design. It requires a parallel path
for the RL and the propagation of light through several substrates
which could change the phase via index of refraction changes. The
factors affecting this phase are discussed in the following subsections.

\subsubsection{Mach-Zehnder Stability}

\label{MZ} The MZ-like interferometer transfers the phase of the
RC transmitted light to the PC transmitted light while blocking PC
transmitted light from reaching the RC. Inside the MZ (see Figure
~\ref{fig:Optics_RL_COB}), the axion travels from MZ3 to MZ4 while
the RL travels an ideally identical path from MZ1 to MZ2 experiencing
identical length changes between the left and right half of the MZ.
The RL also travels from MZ2 to MZ4. A length change of this distance
would change the phase of RL via the first PLL and subsequently the
phase of the PC-transmitted light via the second PLL. However, a parallel
length change between MZ1 and MZ3 changes the phase of RL in the second
beat signal as well canceling the phase change induced by the first
length change. Consequently, these symmetric length changes compensate
each other in first order.

However, this compensation or common-mode rejection is not perfect.
To minimize it, we build the MZ on a ULE base plate. Each mirror is
mounted on a post that is secured inside a hole in a ULE block. The
custom posts are designed such that the relevant surface of each mirror
is centered over the hole~\cite{soham}. This way, the distances
between the surfaces are mostly defined by the ULE block. If we assume
a residual effective coefficient of thermal expansion (CTE) of $10^{-7}/{\rm {K}}$
over each of the $15~{\rm {cm}}$ relevant beam paths, we require
a long term temperature stability of $1~{\rm {K}}$ without common-mode
rejection. With common-mode rejection, we can tolerate much larger
temperature changes and believe this is irrelevant compared to other
noise sources such as changes in the index of refraction.

\subsubsection{Index of refraction changes}

Our substrates are all made from fused silica which has a temperature
dependence of refractive index, $dn/dT$, of $10\,{\rm {ppm/K}}$;
the thermal expansion of the substrate is about a factor 20 smaller
and included in $dn/dT$ for this discussion. Each substrate is $9.5~{\rm {mm}}$
thick, making the optical path length changes for the $35^{\circ}$
tilted substrates about $103~{\rm {nm}/{\rm {K}}}$ and for the normal
incidence cavity optics about $95~{\rm {nm}/{\rm {K}}}$.

One part of the RL will reflect at the front faces of MZ1, MZ2, and
MZ4 and will then propagate through LT2 and the substrate of RC1 before
it combines with the RC transmitted light; we ignore the thin wave
plates in this estimate here. Once the fields are combined, the changes
in the optical path length are common and can be ignored. Consequently,
the thermo-optical change of the phase of the beat between the RL
field and the RC-transmitted field is 
\begin{equation}
\Delta\phi_{RL\_RC}\approx-\frac{2\pi}{\lambda}\left(l_{LT2}\Delta T_{LT2}+l_{RC1}\Delta T_{RC1}\right)\frac{dn}{dT},
\end{equation}
where $l_{X}$ is the optical path through substrate $X$ and $\Delta T_{X}$
its temperature change. The minus sign has been included as we assume
that the LO frequency is larger than the RL frequency which is larger
than the HPL frequency.

The second part of the RL will transmit through MZ1 and MZ3 before
it combines with the PC transmitted light. The PC transmitted light
itself propagates through LT1 before it combines with the RL; the
optical path length changes through the PC mirror substrate will be
discussed next. (We also ignore the thin wave plates in this estimate.)
The thermo-optical changes of the phase of the beat between the RL
field and the PC-transmitted field are then 
\begin{eqnarray}
\Delta\phi_{RL\_PC} & = & -\frac{2\pi}{\lambda}\frac{dn}{dT}\\
 & \times & \left(l_{MZ1}\Delta T_{MZ1}+l_{MZ3}\Delta T_{MZ3}-l_{LT1}\Delta T_{LT1}\right).\nonumber 
\end{eqnarray}
The worst cases are temperature decreases of LT1, LT2, and RC1 combined
with temperature increases of MZ1 and MZ3 resulting in about 500\,nm/K
optical path length changes. However, the optical components are all
in thermal contact through their posts either directly with the aluminum
base plate or with the ULE block which sits on it. Therefore thermal
changes will be, to a large degree, common in the substrates, so:
\begin{eqnarray}
\Delta\theta_{FS} & = & \Delta\phi_{RL\_PC}-\Delta\phi_{RL\_RC}\nonumber \\
 & = & \frac{2\pi}{\lambda}\left(l_{LT1}+l_{LT2}+l_{RC1}-l_{MZ1}-l_{MZ3}\right)\frac{dn}{dT}\Delta T\nonumber \\
 & \approx & \frac{2\pi}{\lambda}l_{RC1}\frac{dn}{dT}\Delta T\approx\frac{2\pi}{\lambda}\frac{100\,{\rm {nm}}}{K}\Delta T.
\end{eqnarray}
We placed a requirement of 0.1\,K absolute temperature stability
on the cleanroom HVAC system during science runs; achieving this stability
will reduce the phase change due to index of refraction changes $\Delta\theta_{FS}$
to $\approx0.06\,{\rm {rad}}$. Following the same logic, index of
refraction changes in the two half-wave plates HW1 and HW2 will add
to the phase drift and controlling the thermal environment will be
critical for the success of ALPS~II.

\subsection{PC cavity mirror substrate}

\label{OPL_Sensing} During science runs, 150\,kW of laser power
circulates inside the PC. Even state-of-the-art coatings are expected
to absorb on the order of 100 to 150\,mW of this power. This absorption
will deform the radius of curvature of the flat mirror on the COB
by approximately 3\,km~\cite{Winkler1991,vinet_reducing_2007} and,
more importantly, raise the temperature of the substrate by a few
degrees. In turn, the optical path length (OPL) through the substrate
will increase by a few hundred nm. During science runs, we will be
sensitive to cavity internal power changes and we will monitor and
potentially stabilize the laser power to minimize the impact on signal
detection. We also integrate into the design a polarization multiplexed
homodyne interferometer, shown in Figure 4, to monitor directly the
OPL changes inside PC2.

\begin{figure}
\begin{centering}
\includegraphics[width=8cm]{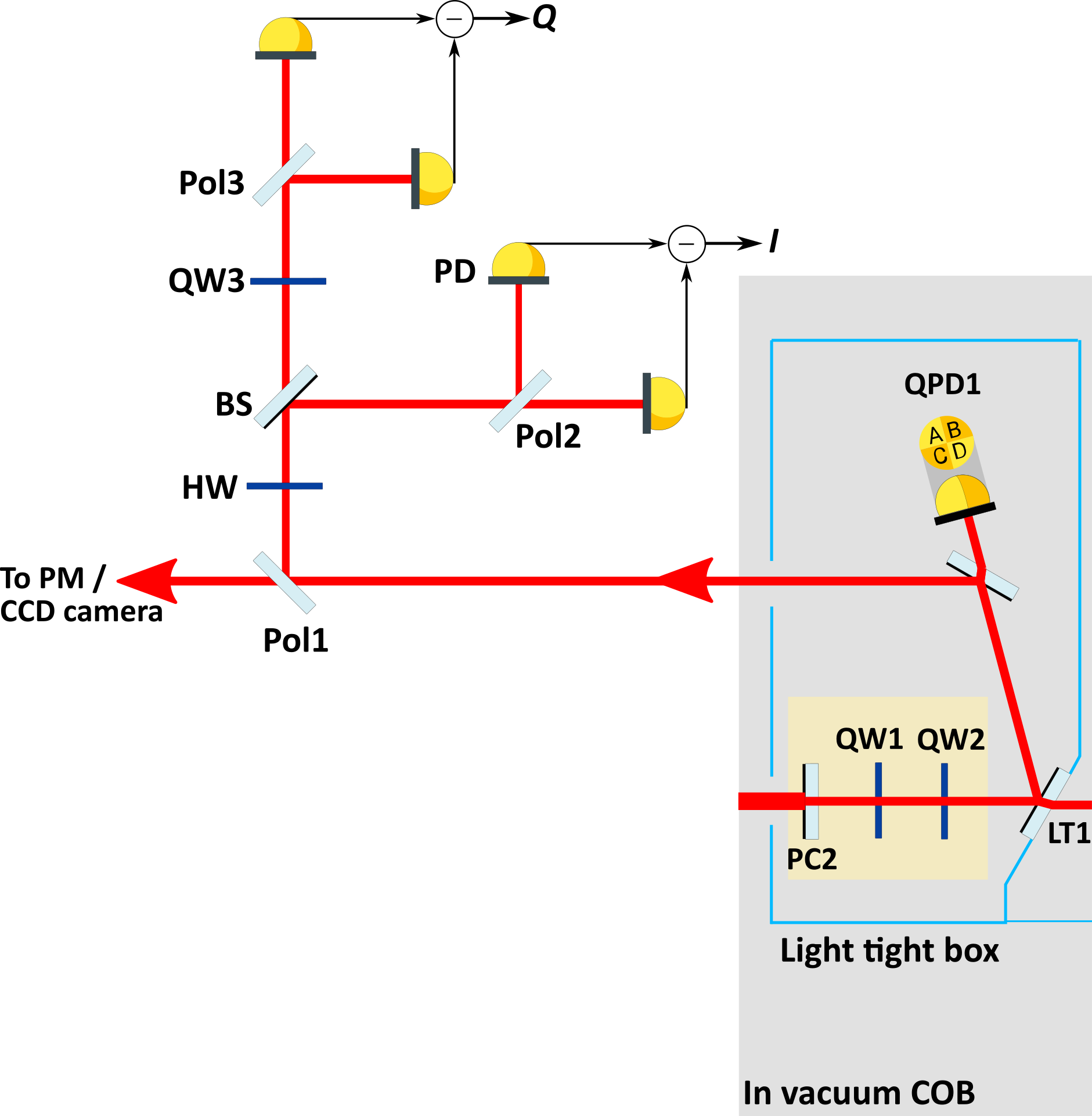} 
\par\end{centering}
\caption{\label{fig:OPL-sensing}Balanced homodyne interferometer used to track
the optical phase changes in the PC transmitted field caused by heating
of the flat cavity mirror substrate. BS: Beam Splitter, COB: Central
Optical Bench, CCD: Charge-Coupled Device, HW: Half-Wave plate, Pol:
Polarizer, PC: Production Cavity, PM: Power Meter, QPD: Quadrant Photodetector,
QW: Quarter-Wave plate, \textit{I}: In-phase signal, \textit{Q}: Quadrature
signal.}
\end{figure}

As mentioned before, we rotate the polarization of the PC output beam
to increase the amount of transmitted light through LT1 for PLL2 and
WFS2. Instead of a single half-wave plate, we use two quarter-wave
plates QW1 and QW2 that are part of the interferometer; the first
turning the s-polarized light into circularly polarized light and
the second turning the circularly polarized light into p-polarized
light. Three of the four surfaces of the quarter-wave plates have
anti-reflection coating while the second surface of QW1 is uncoated
such that 4\% of the circularly polarized light reflects bac towards
the PC. This light passes again through QW1 which turns it into p-polarized
light before it reflects again at the PC. Although the PC is on resonance
with this light, the highly under-coupled PC (seen from this direction)
will reflect most of this light. It then passes through the QW1 and
QW2 which turn it into s-polarized light. The resulting elliptical
polarized light is now a superposition of the initially \textit{s}-polarized
amplitude $E_{s}$ and the initially \textit{p}-polarized probe beam
$E_{p}$, 
\begin{equation}
\vec{E}=E_{s}\hat{e}_{P}+E_{p}e^{i\zeta}\hat{e}_{S}\qquad {\rm with} \quad|E_{p}|^{2}\approx0.04|E_{s}|^{2}
\end{equation}
where $\zeta$ is the phase difference between the two fields: 
\begin{eqnarray}
\zeta & = & \frac{2\pi}{\lambda}\big{(}2\Delta l_{PC2-QW1}\\
 & + & 2(n_{QW1}-1)d_{QW1}+2(n_{PC2}-1)d_{PC2}\big{)}.\nonumber 
\end{eqnarray}
$l_{PC2-QW1}$ is the geometric distance between the HR coated surface
of PC2 and the uncoated surface of QW1. $d_{QW1}$ and $d_{PC2}$
are the two thicknesses of the two optics while $n_{QW1}$ and $n_{PC2}$
are their temperature-dependent indices of refraction. Similar to
the Mach-Zehnder, these two optical components are mounted on a ULE
baseplate which ensures that their distance will not change significantly
with temperature. The main phase change is expected to come from temperature
changes of PC2 which is what the polarimeter is designed to monitor.

The polarimeter itself is placed outside the vacuum chamber and is
shown together with the PC-side of the COB in Figure~\ref{fig:OPL-sensing}.
On leaving the vacuum chamber, this beam first encounters a polarizing
beam splitter Pol1. This beam splitter reflects nearly all \textit{s}-polarized
light and a few percent of \textit{p}-polarized light. This will help
to balance the power in the two polarizations which in turn will improve
the contrast in the final polarimeter signals. This field propagates
through the half-wave plate HW which rotates each of the two polarizations
by $45^{\circ}$, 
\begin{eqnarray}
\vec{E} & = & E_{s}\hat{e}_{p}+E_{p}e^{i\zeta}\hat{e}_{s}\underbrace{\Rightarrow}_{HW-plate}\\
 & \propto & \left(E_{s}+E_{p}e^{i\zeta}\right)\hat{e}_{s}+\left(E_{s}-E_{p}e^{i\zeta}\right)\hat{e}_{p}.\nonumber 
\end{eqnarray}
A polarization-independent 50/50 beam splitter (BS) creates now two
identical beams. The reflected beam is split by a polarizing beam
splitter Pol2, resulting in a power in each of the polarizations of
\begin{eqnarray}
P1_{s}=|E_{s}+E_{p}e^{i\zeta}|^{2}=P_{s}+P_{p}+\sqrt{P_{s}P_{p}}\cos\zeta\\
P1_{p}=|E_{s}-E_{p}e^{i\zeta}|^{2}=P_{s}+P_{p}-\sqrt{P_{s}P_{p}}\cos\zeta~.
\end{eqnarray}
The transmitted beam passes through the quarter-wave plate QW3 which
delays the $(\hat{e}_{s}-\hat{e}_{p})$ polarized light by a quarter-wave
compared to the $(\hat{e}_{s}+\hat{e}_{p})$ polarization, 
\begin{eqnarray}
 & \left(E_{s}+E_{p}e^{i\zeta}\right)\hat{e}_{s}+\left(E_{s}-E_{p}e^{i\zeta}\right)\hat{e}_{p}\underbrace{\Rightarrow}_{QW-plate}\nonumber \\
 & \propto\left(E_{s}+iE_{p}e^{i\zeta}\right)\hat{e}_{s}+\left(E_{s}-iE_{p}e^{i\zeta}\right)\hat{e}_{p}.
\end{eqnarray}
The following polarization beam splitter and the two photodetectors
measure also the power in each polarization state, 
\begin{eqnarray}
P2_{s}=|E_{s}+iE_{p}e^{i\zeta}|^{2}=P_{s}+P_{p}-\sqrt{P_{s}P_{p}}\sin\zeta\\
P2_{p}=|E_{s}-iE_{p}e^{i\zeta}|^{2}=P_{s}+P_{p}+\sqrt{P_{s}P_{p}}\sin\zeta.
\end{eqnarray}
The ratio of the differences between the pairs of signals can be used
to unwrap the phase $\zeta$ between the two original polarized beams,
\begin{eqnarray}
\zeta=\arctan\left(\frac{P2_{p}-P2_{s}}{P1_{s}-P1_{p}}\right).
\end{eqnarray}
Changes in $\zeta$ are proportional to the optical distance between
the uncoated QW1 surface and the HR surface of the PC cavity mirror.

As discussed above, the main cause of phase changes is expected to
be temperature changes in the PC2 substrate: 
\begin{eqnarray}
\Delta\zeta=2kd\frac{dn}{dT}\Delta T.
\end{eqnarray}
Half of this phase change will be encountered by the PC-transmitted
light before it is combined with the RL light and will shift the phase
of the HET and the veto signal. It has to be taken into account during
the final demodulation by adjusting $\theta\rightarrow\theta+\Delta\zeta/2$
accordingly.

When switching from pseudoscalar search runs to scalar searches the
light circulating in the PC will be changed to p-polarization. Consequently,
the relative orientation of quarter-wave plates QW1 and QW2 will be
adjusted such that they compensate each other and no longer change
the polarization in front of LT1, reestablishing the same polarization
states on the rest of the COB as in the pseudoscalar case. In this
case, we will have to add a half-wave plate before Pol1 to balance
the power levels again.

\section{Resonance frequency and alignment optimization and verification \label{sec:Overlap}}
%Should start with a requirement of what we need to achieve here. .

%subsections: Resonance frequency matching,  RC build up of PC-transmitted light (also of RL light). 

The initial alignment process of the COB ensures that the two HR surfaces of the cavity mirrors PC2 and RC1 are parallel within $5\,\mu\rm{rad}$. The parallelism and near-equal thicknesses of the substrates between the HR coatings should guarantee that the PC transmitted light is aligned with the axion mode at the RC to better than $2\,\mu\rm{rad}$ and $250\,\mu\rm{m}$. The initial alignment will also have to ensure that light traveling along the nominal optical axes of the cavities will hit QPD1 and QPD2 withing $500\,\mu\rm{m}$ of their center. Furthermore, the RL will not only be used to transfer the phase information from the RC transmitted field to the PC transmitted field but will also measure the relative alignment between these fields. A careful alignment of the MZ on the COB is required to take advantage of this feature which is limited only by substrate refraction to $2\,\mu\rm{rad}$ and $250\,\mu\rm{m}$~\cite{ALPS-opticspaper}. 

Following the installation of the COB into the vacuum chamber, the laser in each end station will be locked to its cavity and each cavity will be aligned using the signals from the two position sensors QPD1 and QPD2 and feeding back to respective curved mirror pitch and yaw. Following the commissioning of these servo systems, the detectors PLL1 and WFS1 will be aligned to the RC transmitted light. Next, the RL will be phase-locked with an offset frequency of $\Omega_1$ to the LO using a fiber link between the central station and the end station. The RL will then be injected into the COB using the WFS1 signals as a target. Once the RL is aligned, the PLL will be switched over to PLL1 and the alignment will further be optimized using the WFS1 signals. 

Next, the RL will be phase-locked at $\Omega_2$ to the HPL using a second fiber link between the central station and the HPL station. The detectors PLL2, WFS2, PD3, and WFS3 will be aligned to the RL. Once the HPL is locked to the PC, the beat signals should become visible at all four detectors assuming an open shutter. At this stage, it is possible to verify the alignment between the RL and the PC transmitted light at the two sides of the MZ using the contrast in PLL2 and PD3 and the pitch and yaw signals from WFS2 and WFS3. Any misalignment beyond the one recorded during the COB alignment procedure would indicate a change in the COB alignment during installation. Furthermore, it is also possible to switch between PLL1 and PLL2 and evaluate WFS1 and WFS2 signals to measure any added misalignment between the RC and PC. With all these signals in place and knowing that substrate refraction is low enough that the PC transmitted field is a good reference for the axion field, a re-optimization of the alignment of the COB inside the tank is in principle now possible, although we anticipate that it might not be needed. 

During science runs, the alignment of the RL will be actively controlled using WFS1 while the stability of the COB alignment will be monitored using WFS2. Between science runs, the shutter will be opened and WFS3 will be used to measure independently the relative alignment of the two cavities. These checks allow us to quantify and monitor the matching of the axion spatial mode into the RC. 

At this stage, the frequency, length, and alignment sensing and control systems will be optimized to improve lock stretches and to minimize the residual alignment, frequency, and length noise. Once the servo systems provide sufficiently long locking stretches, the difference frequency $\Omega_{sig}$ between the PC and the RC transmitted fields has to be tuned to $13\times \rm{FSR} \pm 1.5\,\rm{Hz}$ to ensure that the regenerated field is resonant inside the RC~\cite{ALPS-opticspaper}. In the HET design, the required difference frequency is determined by phase locking the RL to the RC transmitted field at the expected signal frequency using the beat signal measured at PLL1. The difference frequency is then changed such that the RL frequency is tuned over the RC resonance frequency. The beat signal between the RL transmitted field and the LO reflected from the RC will then be detected by the HET detector in the end station. The signal will be demodulated at the PLL frequency in both quadratures which will allow us to very accurately measure the optimum signal frequency and with it the FSR of the RC. This information will later be used to set the two PLL frequencies $\Omega_1$ and $\Omega_2$ and to calibrate and verify the FSR sensing system. 
\subsection{Calibration of the veto signal}
One of the largest concerns in ALPS is a false positive detection caused for example by stray light. The most likely path for stray light to be detected on the HET detector in the end station in the spatial mode of the local oscillator is to enter the RC through RC2. The HET design includes a veto signal to identify and quantify stray light. This is the beat signal PLL1 between the RC transmitted field and the light reflected at the RC from the COB side. The transmissivity of the input mirror $T_{RC1} = 6.7~\rm{ppm}$ on the COB is much lower than the transmission of the curved mirror on the end table ($T_{RC2} = 107~\rm{ppm}$). We also expect losses between 40 and 60~ppm per round trip ~\cite{ALPS-opticspaper}. Consequently, the RL sees a highly under-coupled cavity for which the ratio between the transmitted power $P_{T}$ on resonance and the reflected power $P_{R}$ will be
\begin{equation}
    \frac{P_{T}}{P_{R}} \approx 0.1
\end{equation} 
So roughly speaking, for every stray light photon that enters the RC through RC1 and is detected at the HET detector, we will detect ten stray light photons with PLL1. In contrast to this, for every 16 photons that were originally generated inside the RC and made it to the HET detector, only one will be detected at PLL1. 

These two ratios provide a veto signal against stray light for the detection of regenerated photons. They will be slightly modified by the losses from the cavity mirrors to the detectors and the efficiencies with which we are able to detect the PLL1 and the HET signal. These losses and efficiencies will be calibrated with the RL laser tuned to resonance and against shot noise of the LO laser field on both detectors \cite{bush_coherent_2019}.

\subsection{Verification of the coupling between PC and RC}
As discussed in Ref.~\cite{ALPS-opticspaper}, the spatial mode of the PC-transmitted field is a good proxy for the spatial mode of the axion field because the optical path between the two cavities has been designed to minimize refraction effects caused by the mirror substrates. The net diffraction of the PC-transmitted field compared to the undisturbed axion mode is expected to be below $2~\mu\mbox{rad}$ while the lateral shift due to differences in the thicknesses of the substrates is expected to be well below $200~\mu\rm{m}$. The resulting spatial mismatch of the modes due to this effect is below 0.4\%, an order of magnitude better than our mode matching goal between the axion mode and the RC. 

This will allow the coupling efficiency and dual resonance between the optical cavities to be verified with the PC transmitted field. With the shutter open, all length, frequency and alignment servos will be engaged and operated at their nominal frequencies. Based on the current knowledge of the injected power, PC cavity losses, and the transmissivities of the various mirrors, about 1.6~pW or $8\times 10^6$\,photons per second will reach PD3, about 100 photons per second will reach the PLL1 or the veto detector and about 10 photons per second will reach the HET detector. The measured ratio between the veto detector and the HET detector has to agree with the ratio measured in the preceding step using the RL to ensure that no systematic errors limit our performance and that we understand reliably the resonant regeneration efficiency of the axion field into the RC.

The long term stability of the entire setup can then be measured with a data run lasting a few days with the shutter open. This run starts with monitoring the amplitude and phase of the beat signal with the PC transmitted light on PD3, the veto signal, and the HET signal on the other side of the RC. It will also include monitoring signals from WFS2 and WFS3 to evaluate drifts in the alignment. Furthermore, the expected lost-lock periods will allow us to test our data analysis method that includes tools to  stitch coherently data runs together using clock signals as a reference.

\section{Stray Light Mitigation \label{sec:Stray}}
% subsections: Power levels of PC-trans light inside IFO, potential of scattered light. Light tight boxes. VETO signals, waveplate switching.
 
ALPS~II must be able to unequivocally identify the regenerated field and distinguish from any background signal at a level equivalent to one regenerated photon a day or $2\times 10^{-24}\,\rm{W}$ for 20~days of data taking. Our goal is to keep the indistinguishable background signal below $2\times 10^{-25}\,\rm{W}$ or one background photon in 10~days. Note that this power level is approximately 30 orders of magnitude lower than the 150kW that are circulating in the PC. 

In the following, we discuss the methods we initially implement to minimize, monitor and eliminate spurious signals in post-processing by:
\begin{itemize}
    \item managing carefully the laser power in conjunction with light-tight boxes and baffles to control and minimize light flux between different areas.
    \item generating signals that can be used as veto signals or even to subtract stray light from the main science signal. 
    \item inverting the contributions from stray light compared to the signal.
\end{itemize}

\subsection{Power management and baffling}

As mentioned in section \ref{sec:Overview}, the mirror coatings were chosen to not only generate the necessary signals but also to minimize stray light and false detections in the HET detector. We defined four different areas on and around the COB (see Figure~\ref{fig:Optics_RL_COB}):
\begin{enumerate}
    \item the PC area includes the PC cavity end mirror PC2, the two quarter-wave plates QW1 and QW2, the position sensor QPD1 on the COB and the OPL sensing system, and the power monitoring and CCD camera on the PC side of the COB.
    \item the RC area includes the RC cavity end mirror RC1, the half-wave plate HW2, the position sensor QPD2 on the COB and the PLL1/Veto channel and the WFS1 sensors on the RC side of the COB. 
    \item the PC/MZ area includes the half-wave plate HW1 on the COB and the PLL2 and WFS2 sensors on the PC side of the COB.
    \item the RC/MZ area includes the MZ  like-interferometer mirrors, the shutter S1 and the beam dumps BD1 and BD2 on the COB, and the PD3 and WFS3 sensors on the RC side of the COB.
\end{enumerate}

\begin{figure*}[t]
\begin{centering}
\includegraphics[width=16cm]{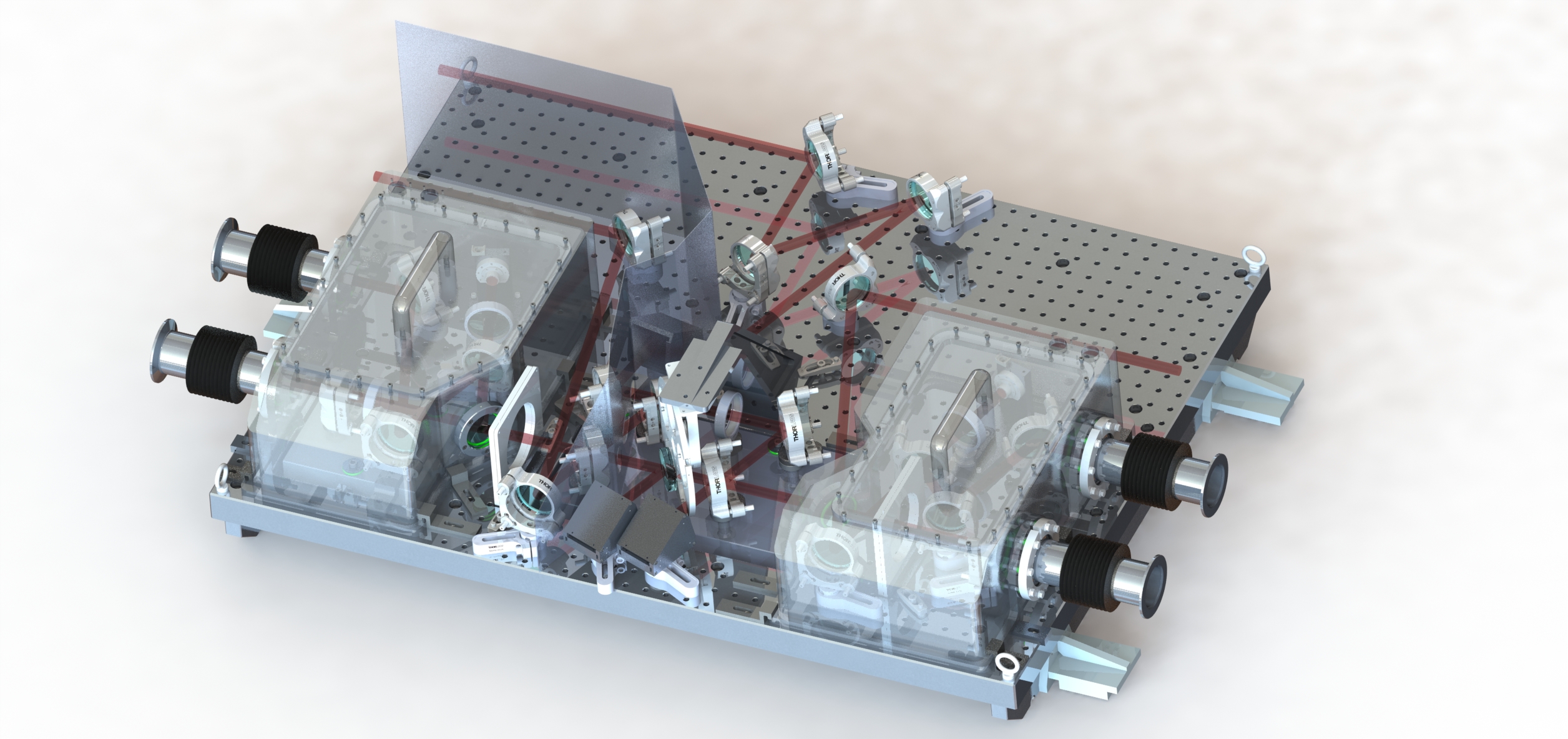}
\par\end{centering}
\caption{\label{fig:COB_joe} Rendering of the central optical bench. The PC area on the left side of the picture is contained within light-tight walls and light can only escape through the cavity port, the OPL sensing port and LT1. A similar light-tight box encompasses the RC area which connects only to the RC, the PLL1/WFS1 detection area and to the rest of the COB via LT2. The wall across the MZ area has not yet been finalized. }
\end{figure*}

Inside the vacuum tank, the RC and the PC areas are enclosed by light-tight aluminum walls. Each of these two chambers includes a light-tight lid and is connected inside the vacuum tank via a vacuum bellow to a Brewster-angle viewport. The reflected light from the viewport is incident on a black-glass beam dump to minimize further back reflection. Outside the vacuum tank, each beam is enclosed by a tube until it reaches the optical table. On the optical table, the distinct areas are separated from each other with a  system of walls to  minimize the chances for PC-transmitted light to re-enter the vacuum chamber; for example, it could come in together with the  RL light. The only optical interfaces between the PC and the PC/MZ as well as between the RC and the RC/MZ areas are the HR surfaces of LT1 and LT2, respectively. Furthermore, the PC/MZ area is separated from the RC/MZ area with baffles to limit the optical leakage of light passing through MZ3. Again, both areas are also enclosed on the optical tables to minimize chances of PC transmitted light to scatter into the RC area. The light blue lines in Figure~\ref{fig:Optics_RL_COB} indicate the light-tight boxes and baffles in the optical layout while Figure~5 shows a near-complete rendering of the central optical bench.

The light that passes through LT1 and enters the PC/MZ area has to undergo at least two scatter processes out and back into the beam before it reaches LT2. Assuming that about 10\,ppm per process scatter light into the acceptance angle leading towards the RC and that up to 10\% of the light is changing polarization during each scattering process, we estimated that less than $10^{-26}\,\rm{W}$ of \textit{s}-polarized light and less than $10^{-25}\,\rm{W}$ of \textit{p}-polarized light will reach the RC mirror in the spatial mode of the RC. Note that good optics and beam dumps under typical operational angles have scatterring values between $10^{-5}\,\rm{sr}^{-1}$ and $10^{-6}\,\rm{sr}^{-1}$~\cite{Vander-Hyde} which, scaled by the acceptance angles in our setup, means that with our above assumption we grossly overestimate the amount of scattered light present. 

However, while the vacuum viewports are all set at Brewster's angle with the residual reflection being directed into a black glass beam dump, the optical components including the photodetectors in the PC/MZ area and in the RL injection area (nominally part of the RC/MZ area) outside the vacuum system will scatter light back into the vacuum system. Here we will use curved mirrors instead of lenses for mode matching and tilt all other components to avoid direct back reflection into the vacuum system. This estimation assumes that scattered light does not bypass any of the mirrors which emphasizes the importance of the light-tight walls and baffles. 

\subsection{Veto Signals}
Despite all the efforts described in the preceding section, hunting and eliminating stray light will certainly be a significant part of the commissioning effort. Here the detector PD3 plays a key role. LT2 at the interface to the RC area reflects nearly all scattered light and directs it towards PD3 where it will form a beat signal with the RL. The rate of scattered photons on PD3 in the correct mode will be around a million times larger than on the RC side of LT2. Thus, it {\it only} has to be suppressed below about one photon per second to meet our requirements. This process will significantly reduce the commissioning time as we do not have to search for scattered light at the regenerated photon rate. 

Light that passes through LT2 and is not in the spatial mode of the RL will have to scatter inside the RC area into the eigenmode of the RC before it will be able to pass through the RC. However, as seen from the COB, the RC is an under-coupled cavity and even on resonance 90\% of the scattered photons will be detected in reflection by the veto detector and only 10\% will make it to the HET detector. In contrast, regenerated photons inside the RC are 20 times more likely to be detected by the HET detector than the veto detector. There is thus the possibility to subtract scattered light from the HET signal if needed.

\subsection{Changing the phase of the signal}
Scattered light that enters the RC from the COB has to pass through the motorized rotating half-wave plate HW2 in front of the RC. During science runs, the main axes of the half-wave plate will be aligned with the {\textit s} and {\textit p} polarization axes, where it does not change the polarization of the laser fields. However, after some time the half-wave plate will be rotated by $90^{\circ}$. The rotation of the half-wave plate will delay the phase of the RL by $\pi$ which will be compensated by inverting the actuation or error signal in PLL1. These two steps combined will ensure that the phase of the regenerated field is not changing. However, all scattered light that passes through the half-wave plate will be phase-shifted which leads to an inversion of its contribution to the veto and the heterodyne signal. As a result, scattered light will destructively interfere and, if repeated with the appropriate periodicity, average away while the signal continues to constructively interfere and builds up. This method will be tested and calibrated with the shutter open and used together with the common-mode rejection from the veto channel during shutter-closed data taking periods to reject further scattered light.

\section{Summary}

ALPS II is a challenging experiment, requiring an optical design that  ensures that the detection efficiency is well understood and calibrated and that as much as possible prevents scattered light from creating false signals. On the one hand, the heterodyne detection system (HET)  described here takes advantage of the coherent nature of the interaction. The coherence allows us to filter out virtually all background photons because only light at exactly the right frequency will generate a signal. On the other hand, this coherence places significantly more stringent requirements on the long-term stability of the setup compared to, for example, a simple photon counting scheme. A Mach-Zehnder-like interferometer setup was developed to superimpose metrology beams, the reference laser beams, with the cavity transmitted light fields. The resulting beat signals are used not only to maintain a 90\% coupling efficiency (in power) of the axion field into the regeneration cavity but also to monitor the long term alignment of the setup.

Our design uses ultra-stable mirror mounts on ultra-low expansion (ULE) base plates to ensure the necessary stability. The near symmetric design also reduces thermo-optical changes from most components while we also include a polarization multiplexed homodyne interferometer to measure optical path length changes in the laser-heated production cavity mirror. The design also responds to the need to suppress stray light as much as possible and provides signals to measure the remaining scattered light at ports that have a much higher susceptibility to it than the HET readout port. The high signal-to-noise ratios in these ``scattered light meters'' will greatly simplify the commissioning of the experiment and potentially even provide options to subtract it from the signal. 

Based on the design parameters, it should be possible to place an upper limit on the axion to two-photon coupling of $g_{a\gamma\gamma} < 2\times 10^{-11}\,\rm{GeV}^{-1}$ within 20 days of integration time unless a detection derails our plans.

\section*{Acknowledgments}
The Authors would like to thank the members of the ALPS Collaboration. We acknowledge the support of the National Science Foundation (Grant No. 1802006), of the Heising-Simons Foundation (Grant No. 2015-154 and 2020-1841) and of  the Deutsche Forschungsgemeinschaft through project grant WI 1643/2-1.

\bibliographystyle{unsrt}
\bibliography{HET_PRD}


\end{document}